\documentstyle[prl,aps]{revtex}
\draft
\input epsf
\input{epsf.sty}
\begin{document}

\twocolumn[\hsize\textwidth\columnwidth\hsize\csname
@twocolumnfalse\endcsname

\title{
Lower Critical Field at Odds with A S-Wave Superconductivity in The New Superconductor $MgB_2$
}

\author{
S. L. Li, H. H. Wen\cite{responce}, Z. W. Zhao, Y. M. Ni, Z. A. Ren, G. C. Che, H. P. Yang, Z. Y. Liu and Z. X. Zhao}

\address{
National Laboratory for Superconductivity,
Institute of Physics and Center for Condensed Matter Physics,
Chinese Academy of Sciences, P.O. Box 603, Beijing 100080, China
}

\maketitle

\begin{abstract}
The lower critical field $H_{c1}$ has been carefully measured on a well shaped cylindrical sample of the new superconductor $MgB_2$ fabricated by high pressure synthesis. The penetration depth $\lambda$ is calculated from the $H_{c1}$ data. It is found that a linear relation of $H_{c1}(T)$ appears in whole temperature region below $T_c$. Furthermore a finite slope of $dH_{c1}/dT$ and $d\lambda(T)/dT$ remains down to the lowest temperature ( 2 K ). These are inconsistent with the expectation for a widely thought s-wave superconductivity in $MgB_2$. 

\end{abstract}

\pacs{74.25.Bt, 74.20.Mn, 74.72.Yg}

]
Recently discovered new superconductor $MgB_2$ generates a new round of chasing in the field of superconductivity\cite{akimitsu}. Several thermodynamic parameters have already been derived, such
as the upper critical field $H_{c2}$ = 15 - 20.4 T\cite{bud1,canfield,takano,finnemore}, the Ginzburg-Landau parameter $\kappa \approx$ 26\cite{finnemore}, and the superconducting critical current density $j_c\approx 1.05 \times 10^5 A/cm^2$ at 15 K and 1 T\cite{wen}. Recently, Hall effect measurement has been done showing that the charge carriers are holes with very high density $n_s = 1.5\times10^{23}cm^{-3}$\cite{kang}. Measurement on Seebeck coefficient\cite{lorenz} signals also positive charges in $MgB_2$. Theoretically there are two major diverse pictures accounting for the superconductivity\cite{kortus,hirsch}. Based on band calculations, Kortus et al.\cite{kortus} proposed that the superconductivity is resulted from the strong electron-phonon interaction and the high phonon frequency associated with the light B element. This picture is supported by the recent isotope measurement\cite{bud2} on $^{10}B$ and $^{11}B$ yielding an isotope exponent $ \alpha \approx$ 0.26. In the second picture, Hirsch suggested\cite{hirsch} an alternative explanation which conjectures the same mechanism of superconductivity in $MgB_2$ as in high temperature cuprate superconductors ( HTS ). Recently high pressure measurement shows that the superconducting transition temperature decreases linearly at a rate of $dT_c / dP$ = -1.6 K/GPa\cite{lorenz}, being in contrast to that predicted by Hirsch's theory. In order to know the superconducting mechanism, to probe the nature of superconducting gap and symmetry seems to be very important. Recent data from specific heat\cite{kremer}, tunneling spectroscopy\cite{rubio,karapetrov,sharoni}and NMR study\cite{kotegawa} commonly suggest a phonon-mediated BCS picture with the s-wave superconducting energy gap $\Delta$ ranging from 2 to 8 meV corresponding to the coupling strength from weak to strong. For a s-wave superconductor, it needs a minimum energy to excite the charges from the superconducting condensate to the excited states. Thus one can expect an exponentially activated temperature dependence in low temperature region for many thermodynamic parameters, such as the lower critical field $H_{c1}$ and penetration depth $\lambda$, etc.. Therefore one should observe a vanishing slope $dH_{c1}(T)/dT$ = 0 and $d\lambda(T)/dT$ = 0 in low temperature region. In this Letter, we present a precise determination of the lower critical field $H_{c1}$ in whole temperature region. Our data are clearly at odds with the prediction for a s-wave superconductivity in $MgB_2$.  

Samples of $MgB_2$ investigated here were fabricated by high pressure synthesis ( P = 6 GPa at 950 $^{\circ}$C for 0.5 hour) described in detail elsewhere\cite{ren}. The X-ray diffraction ( XRD ) analysis shows that it is nearly in a single phase with the second phase less than 1 wt.\%. Our samples are very dense and hard with a metallic shiny surface after polishing. They look also like the hard $YBa_2Cu_3O_7$ bulks grown by melt-textured-growth technique, showing probably the trivial influence of possible weak links between grains ( if any ). Fig.1 shows that the onset $T_c^{onset}$ is 40 K and the transition width is less than 1 K. The inset of Fig. 1 shows a very sharp diamagnetic transition and a perfect diamagnetic signal, in consistent with the resistive result. To minimize the demagnetization factor, the sample was carefully cut and ground to a cylinder with a diameter of 1.58 mm and
\begin{figure}[h]
	\vspace{10pt}
    \centerline{\epsfxsize 8cm \epsfbox{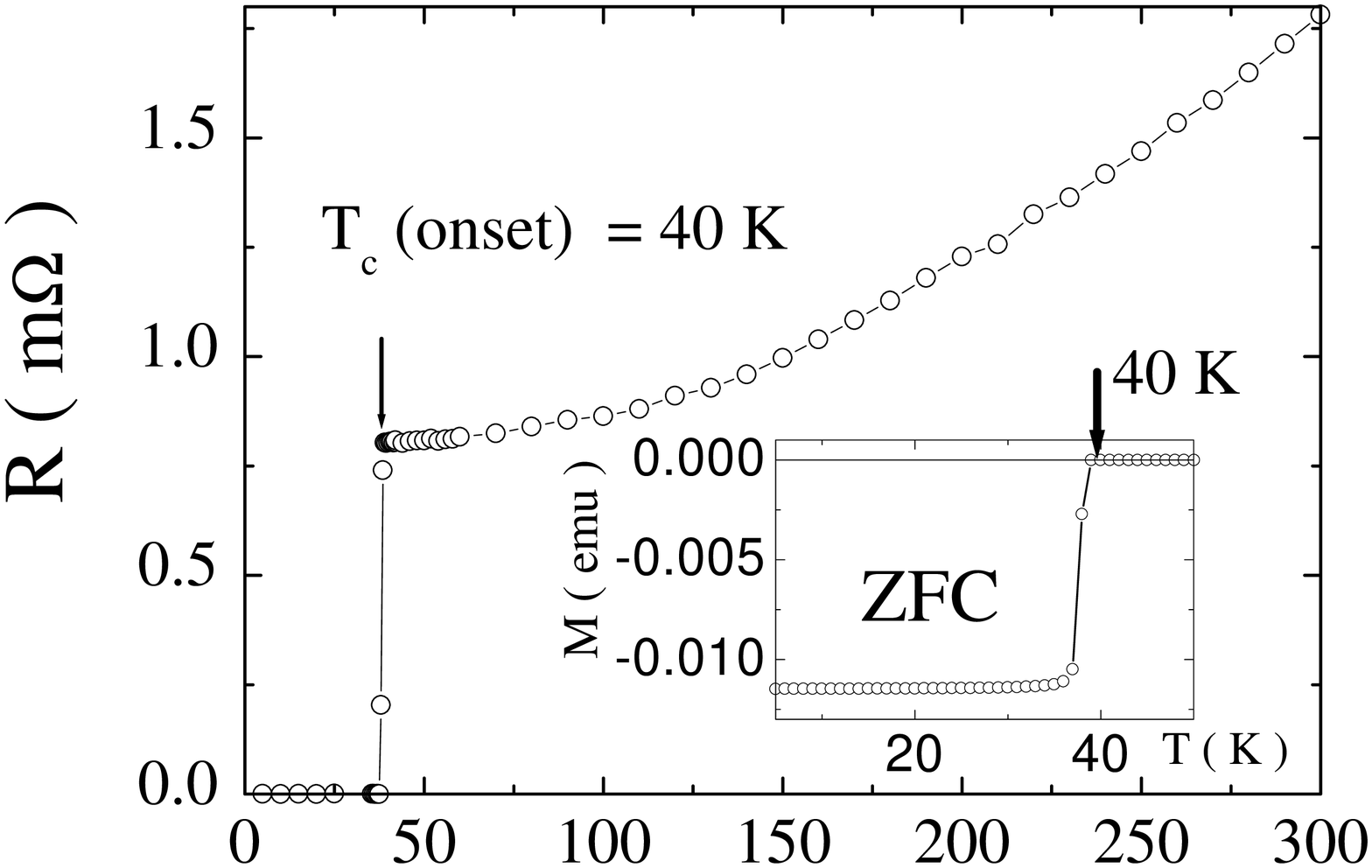}}
    \vspace{10pt}
\caption{The resistance measurement of $MgB_2$. The onset $T_c$ is 40K, and the transition width is less than 1 K. The inset gives the ZFC diamagnetization at 10 Oe, which shows a perfect diamagnetic signal.}
\label{fig:Fig1}
\end{figure}
\noindent  length 3.50 mm. The DC magnetic measurements were carried out by a vibrating sample magnetometer ( VSM, Oxford 3001 ) 
and a superconducting quantum interference device ( SQUID, Quantum Design MPMS 5.5 T ). During the magnetic measurements, the sample axis is always aligned along the field direction so that the demagnetization effect can be neglected. After zero-field-cooled from 50 K to a desired temperature, the magnetic field is applied slowly up to 900 Oe, which is much higher than H$_{c1}$. Fig. 2(a) and 2(b) show the DC magnetization curves obtained by VSM and SQUID respectively. All curves show clearly a common linear dependence of the magnetization on field caused by Meissner effect at low fields. 

The value of H$_{c1}$ is determined by examining the point of departure from linearity on the initial slope of the magnetization curve. To do this, a linear fit with points between 10 Oe and 100 Oe at 2 K is made. This line passes also through the data points measured at other temperatures in low field region. The fitted linear line is called 
\begin{figure}[h]
	\vspace{10pt}
    \centerline{\epsfxsize 8cm \epsfbox{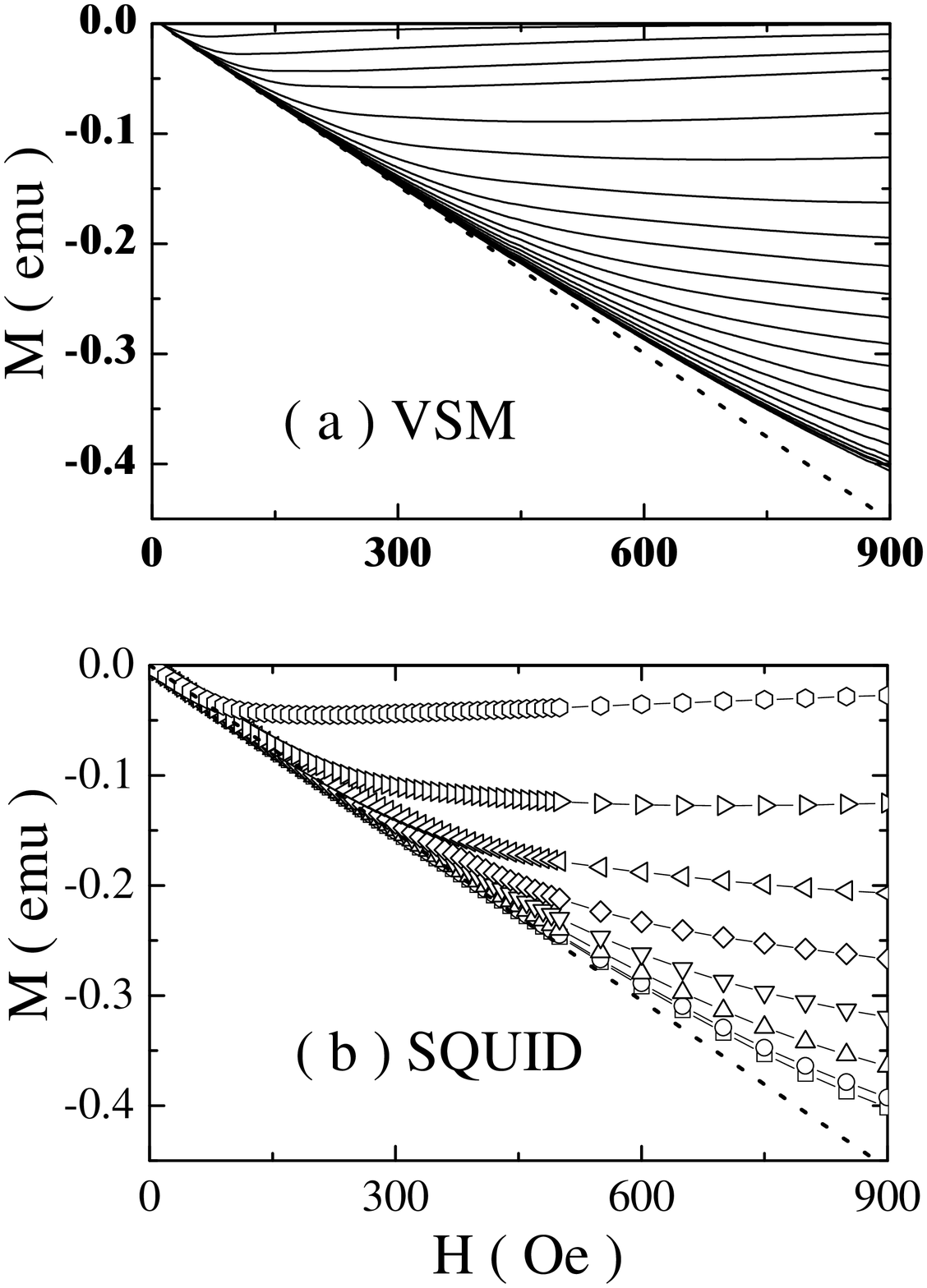}}
    \vspace{10pt}
\caption{The magnetization curve M(H) measured by using (a) VSM (b) SQUID. The dotted line is the " Meissner line " defined in the text. The temperature is selected as (a) 2-6 K step 1 K, 8-32 K step 2 K, 34-38 K step 1 K and (b) 2 K, 5-35 K step 5 K. It is very clear that the initial slope of all the curves is the same.
}
\label{fig:Fig2}
\end{figure}
\noindent the " Meissner line " ( ML ), as the dotted line shown in Fig. 2, which represents the magnetization curve of Meissner state. The results of subtracting this ML from magnetization curves of SQUID measurements are plotted in Fig. 3 in a logarithmic scale. It is important to note that one should avoid to use a linear scale plot of $\Delta$M vs. H to determine $H_{c1}$ since that will induce great uncertainty.  The data from VSM bear close resemblance, thus they are not shown here. The value of H$_{c1}$ is easily obtained by choosing a proper criterion of $\Delta$M. In Fig. 4(a), we plot the H$_{c1}$ acquired by using a criterion of $\Delta M = 1 \times 10^{-3}  emu$. The H$_{c1}$ measured by VSM and SQUID coincide in the same line, showing that our results are independent on measuring device and technique. Extrapolating the H$_{c1}$ curve to zero temperature, we get H$_{c1}$(0) $\approx$ 295 Oe, being close to 320 Oe found by Takano et al.\cite{takano} A striking feature here is that the curve $H_{c1}(T)$ shows a good linear behavior in whole temperature region as shown by the straight line in Fig. 4(a).

To obtain the penetration depth $\lambda$, we use the expression
\begin{equation}
H_{c1}=\frac{\rm \Phi_{0}}{\rm 4\pi \lambda ^2}ln\kappa,
\label{eq1}
\end{equation}
where $\Phi_{0}=hc/2e\approx 2\times 10^{-7} G\cdot cm^2$ is the flux quantum, and $\kappa \approx$ 26\cite{finnemore} is the Ginzburg-Landau parameter. Using H$_{c1}$(0) = 295 Oe obtained above, we find $\lambda$(0) $\approx$ 1325 $\AA$, being close to $\lambda$(0) = 1400 $\AA$ found by Finnemore et al.\cite{finnemore}. The deviation of the penetration depth from its zero temperature value, $\Delta\lambda = \lambda(T)-\lambda(0)$, is shown in Fig. 4(b). It is remarkable that the dependence of $\Delta\lambda$ on temperature below 20K ($T_c/2$) is almost linear, i.e., $\Delta\lambda \propto T$, which is more explict in the inset. A rough estimate gives the slope $d\lambda/dT \approx 28 \AA/K$. 
\begin{figure}[h]
	\vspace{10pt}
    \centerline{\epsfxsize 8cm \epsfbox{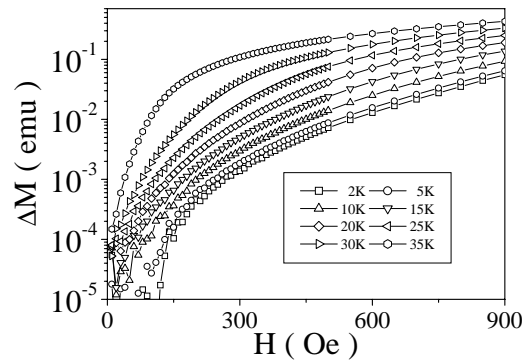}}
    \vspace{10pt}
\caption{The difference between the linear line and the M(H) curves measured on the SQUID. $\Delta M$ is shown in logarithmic scale. All curves show a fast drop when the real $H_{c1}$ is approached. By using a proper criterion, e.g., $10^{-3}$emu, $H_{c1}$ is easily obtained.
}
\label{fig:Fig3}
\end{figure}

In order to know whether this linear behavior is induced by the uncertainty in using the criterion for $\Delta $M, we have used different criterions, such as $1 \times 10^{-4} emu$, for the determination of $H_{c1}$, but the linear dependence is not changed. This linear dependence conflicts the prediction for a s-wave superconductivity that supported by some expriments mentioned in the first paragraph of this paper. 

For a S-wave superconductor, the finite energy gap manifests itself with an exponentially activated temperature dependence of many thermodynamic properties. In the s-wave BCS theory, $\Delta\lambda$ is given by\cite{muhlschlegel},
\begin{equation}
\frac{\rm \Delta\lambda(T)}{\rm \lambda(0)}\sim3.3(\frac{\rm T_c}{\rm T})^{\frac{\rm 1}{\rm 2}}exp(-\frac{\rm \Delta}{\rm k_BT_c}\frac{\rm T_c}{\rm T}),
\label{eq2}
\end{equation}
\noindent where $\Delta$ is the energy gap. The exponential term causes \\
\begin{figure}[h]
	\vspace{10pt}
    \centerline{\epsfxsize 8cm \epsfbox{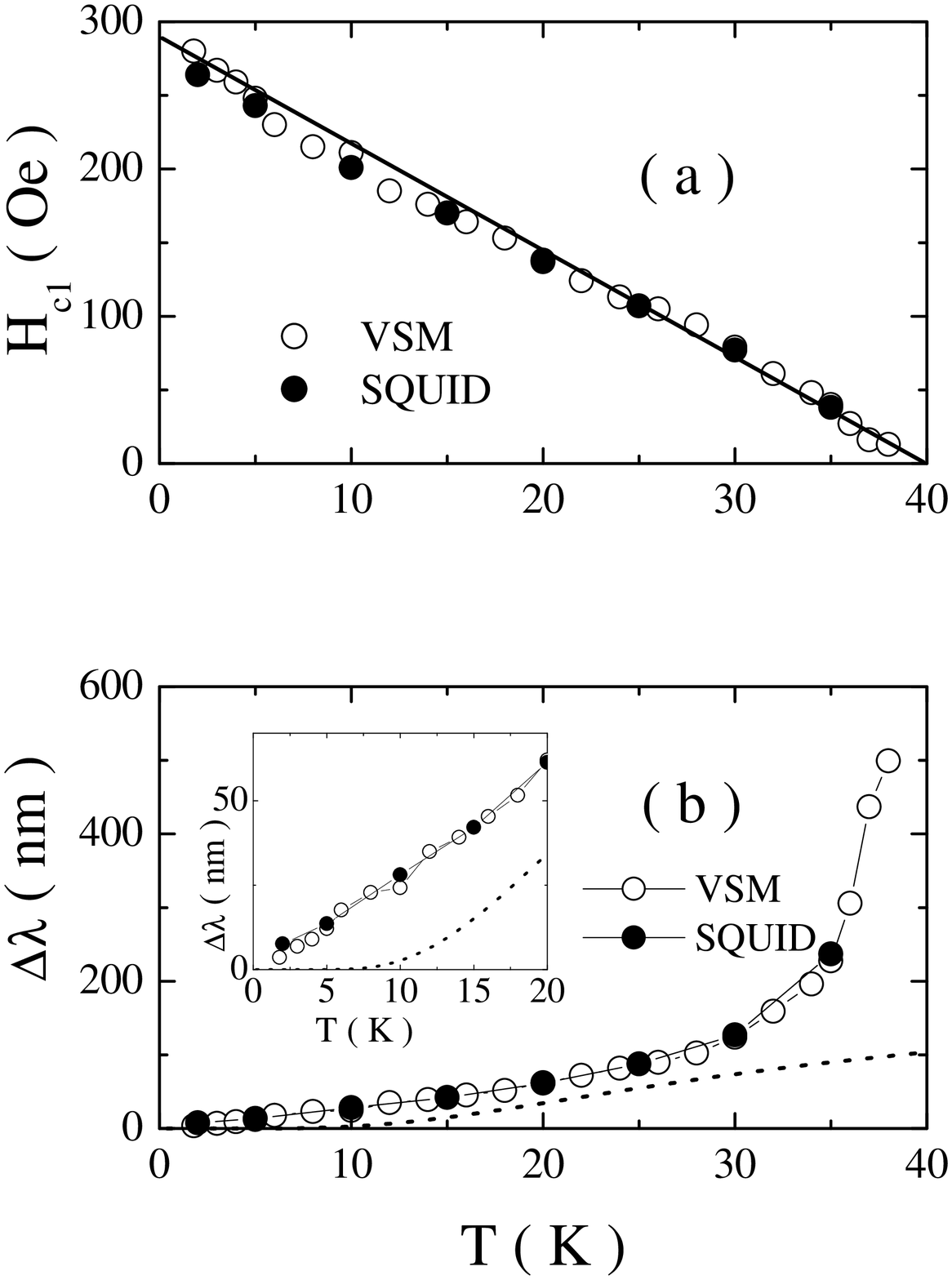}}
    \vspace{10pt}
\caption{The temperature dependence of the (a) $H_{c1}$, (b) $\Delta\lambda$. The solid circles and open circles represent data obtained by SQUID and VSM respectively. The low temperature part is enlarged in the inset of (b), which shows a clear linear T dependence of $\Delta\lambda$. The dotted line in (b) and its inset is a theoretical calculation by using Eq. (2). Because there is no consistent value of energy gap $\Delta$ yet, we choose a intermediate value, $\Delta$ = 5 meV. It should be noted here that the value of $\Delta$ only affects the magnitude of $\Delta\lambda$ and does not change the low temperature behavior of $\Delta\lambda$.
}
\label{fig:Fig4}
\end{figure}

\noindent \\the very weak temperature dependence of $\Delta\lambda(T)$, i.e., $d\lambda(T)/dT \approx 0$, as shown by the dotted line in the inset of Fig. 4(b). It is easy to see that the linear dependence of $\Delta\lambda(T)$ versus T shown in Fig. 4(b) and its inset is clearly different from the prediction by the s-wave BCS theory. Even for a strong coupling superconductor with probably an anisotropic energy gap, it is still hard to understand the finite slope of $H_{c1}(T)$ and $\lambda(T)$ in low temperature region. There was an argument contributing this linearity to the classical superconducting phase fluctuation\cite{roddick}. This possibility can be however ruled out by the observation of very weak fluctuation effect in $MgB_2$ characterized by the non-rounded onset superconducting transitions measured at different fields. Another argument would be that our samples contain still grains and thus weak links. This is basically correct, but as mentioned already, our samples are prepared by high pressure synthesis with very high density and hardness together with rather high critical current density, therefore the couplings between grains ( if any ) are very strong. Furthermore, the zero temperature penetration depth $\lambda(0)$ determined from our high pressure synthesized cylindrical sample is very close to that found by Finnemore et al. based on a consideration on the intrinsic parameters of $MgB_2$. This may manifest that the linear $H_{c1}(T)$ determined in our present work is an intrinsic property.

This linear behavior is similar to that in HTS. In cuprate superconductors, it has been ascribed to the d-wave symmetry\cite{hardy,bonn} of energy gap based on the consideration as follows. In a pure d-wave superconductor, the energy gap along the node directions ( $k_x = \pm  k_y$ ) vanishes and the spectrum $N_s(E)$ of quasiparticle excitations in the superconducting phase is gapless, which results in the linear dependence of $N_s(E)$ on E at low excitation energies. A finite temperature will excite certain quasiparticles leading to a linear dropping of the superfluid density $\rho_s$ with temperature. As to other p-wave superconductors, $\Delta\lambda$ also shows non-exponential dependence on T because of the nodes of order parameter\cite{annett}. The linear $H_{c1}(T)$ measured in $MgB_2$ may not imply a d-wave or p-wave superconductivity, but calls certainly for a new understanding. For example, according to the theory by Hirsch et al.\cite{hirsch}, it appears to be the same " hole superconductivity " in HTS and $MgB_2$. Further investigation may lead to an answer to the question whether the finite slope of $H_{c2}(T)$ and $\Delta \lambda (T)$ at low temperatures for HTS and $MgB_2$ are due to the same cause. In addition, the penetration depth measurement by other techniques, such as $\mu SR $, microwave and a.c. susceptibility is very desirable in order to have a comparison to our data. 

In conclusion, we have measured the M-H curve of a cylindrical $MgB_2$ sample and obtained its lower critical field $H_{c1}(T)$. A striking linear curve $H_{c1}(T)$ has been observed in the whole temperature region. Further calculation of the penetration depth shows also a finite slope in the low temperature limit. All these are at odds with the expectations for a s-wave superconductivity. The reason is unknown but warrants certainly further investigation in order to get a final consensus of the order parameter symmetry of $MgB_2$.

\acknowledgements
This work is supported by the National Science Foundation of China (NSFC 19825111) and the Ministry of Science and Technology of China ( project: NKBRSF-G1999064602 ).

\end{document}